# Depth estimation of tumor invasion in early gastric cancer using scattering of circularly polarized light: Monte Carlo Simulation study

Nozomi Nishizawa[1]*, and Takahiro Kuchimaru[2]

[1] *Laboratory for Future Interdisciplinary Research and Technology, Tokyo Institute of Technology, Yokohama 226-8503, Japan.*

[2] *Center for Molecular Medicine, Jichi Medical University, Tochigi 329-0498, Japan.*

E-mail: nishizawa.n.ab@m.titech.ac.jp

**ABSTRACT**

Quantitative depth estimation of tumor invasion in early gastric cancer by scattering of circularly polarized light is computationally investigated using the Monte Carlo method. Using the optical parameters of the human stomach wall and its carcinoma, the intensity and circular polarization of light scattered from pseudo-healthy and cancerous tissues were calculated over a wide spectral range. Large differences in the circular polarization with opposite signs, together with 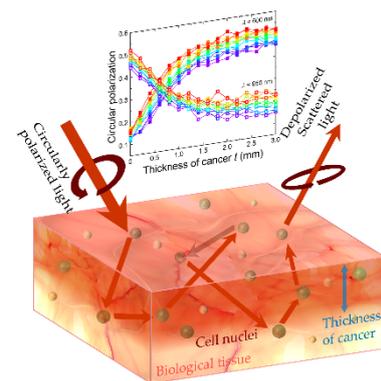 the large intensity, are obtained at wavelengths 600 nm and 950 nm. At these two wavelengths, the sampling depth of the biological tissues can be modulated by tuning the detection angle. In bi-layered pseudo-tissues with a cancerous layer on a healthy layer and vice versa, the degree of circular polarization of scattered light shows systematic changes depending on the thickness and depth of the cancerous layer, which indicates the feasibility of *in vivo* quantitative estimation of cancer progression in early gastric cancer.

# 1. Introduction

Recent developments in diagnostics and treatments are one of the major causes for the gradual reduction in mortality rate due to gastric cancer in the past few decades [1]. Nevertheless, gastric cancer remains a serious health issue, especially in East Asian countries. Above all, for Japanese males, the age-standardized incidence rate of stomach cancer per 100,000 is extremely high, which is probably because of the high infection rate of chronic *Helicobacter pylori* and large intake of salted foods.

Early detection of cancer is the most crucial strategy for increasing the chance of effective treatment and reducing mortality due to cancer. Accurate diagnostic results regarding the tissue types, locations, distributions, and progress degrees of the cancer are required for the appropriate treatment. Gastric cancers have been classified according to the TNM staging system, where the three key parameters are T (size or direct extent of the primary tumor), N (degree of spread to regional lymph nodes), and M (presence of distant metastasis). Early-stage cancers have been identified almost only with T stage classification (Figure 1) [2-4]. The cancers categorized in the Tis and T1 stages rarely spread to a lymph node. It can be removed without sacrificing the entire organ by using endoscopic surgical procedures, such as endoscopic submucosal dissection (ESD) [5-7]. Approximately 20 to 30 % of cancers in T2 stages spread to lymph nodes, which have been treated by surgical treatment with lymph node dissections. Therefore, accurate tumor invasion depth in gastric cancer is of primary importance in determining the therapeutic approach. Recently, image-enhanced endoscopy (IEE), represented by narrow-band imaging (NBI) [8], which is used for the diagnosis of tissue types and degree of spread in cancer has drastically improved the quantitative and qualitative diagnosis of gastrointestinal cancers. Magnifying endoscopy with NBI (ME-NBI) has been used for determining the invasion depth by observing the intrapapillary capillary loops for esophageal tumors and pit patterns for colon tumors [9]. For these tumors, ME-NBI has obtained a high diagnosis rate. In contrast, for the depth estimation of tumor invasion for gastric cancers, magnifying endoscopy with indigo carmine dye contrast or endoscopic ultrasonography (EUS) has been used. The correlation between the invasion depth and abnormal blood vessels or mesh patterns has been reported in recent years [10, 11]. However, only a few cases were analyzed for these studies. Methods to determine the invasion depth in gastric cancer have not yet been established, necessitating a breakthrough from a different technological viewpoint.

One of the techniques is the optical biological observation using light polarization [12]. Scattering of incident linear polarized light (LPL) provides structural information of biological tissues. This is carried out by analyzing the degree of birefringence and depolarization from the supplying images that reflect the anisotropy of the tissues [13-17]. However, in a turbid medium like biological tissues, LPL is readily lost by multiple scattering, therefore, it provides poor information from deep regions of tissues. In contrast, circularly polarized light (CPL) has comparatively more endurance against multiple scattering than LPL [18, 19]; its polarization survives even after a large number of scattering events in biological tissues with a thickness of several millimeters. Utilizing this characteristic of CPL, Meglinski *et al.* [20] pioneered the application of CPL for optical cancer detection by mapping the polarization-dependent optical properties of tissues on the Poincaré sphere. They verified the use of the Stokes vector of backscattered light from tissues for non-invasive optical tissue biopsy. Kunnen *et al.* [21] reported that when CPL impinges on a human lung tissue, the polarization states of the scattered light show clear differences between normal and tumor tissues. These findings were interpreted as follow: the CPL beams impinged on a biological tissue are scattered multiple times by cell nuclei and gradually depolarized. In the Mie regime, where the scatterers (cell nuclei) are larger than the wavelength, the depolarization sensitively depends on the size of the scatterers [22]. Therefore, the resultant degree of circular polarization (DOCP) of the scattered light provides structural characteristics, such as the size, anisotropy, distribution, and density of the cell nucleus. In cancerous tissue, the cell nuclei are enlarged due to the abnormal growth of cancer, which can be detected by the difference in the DOCP of the normal and tumor tissues. Following these reports, the polarimetry technique for tissue observation using the polarization of light, including CPL, has been extensively studied [23-30]. Recently, the polarimetry technique has been applied for grading colon cancer [31], Alzheimer's disease [32], and early-stage breast cancer [33].

Following these studies, we also have investigated the scattering phenomena of incident CPL in biotissues for cancer detection. Before then, we have studied CPL-emitting diodes, also called spin-LEDs, and we have achieved fully polarized CPL emission at room temperature [34], electrical switching of CPL helicity at high speed [36], and detection of CPL at room temperature [37, 38]. Based on these achievements, we proposed CPL scattering technique for *in vivo* diagnosis of gastrointestinal cancer by combining the cancer detection technique with CPL scattering which Meglinski *et al.* developed and spin-LEDs which we have developed. CPL cannot be transferred through a tortuous optical fiber, while maintaining its polarization. However, the spin-LED can directly emit, control, and detect CPL even at spatially restricted places. The spin-LEDs integrated at the tip of an endoscope allow *in vivo* cancer detection using the CPL scattering technique. This technique does not require any staining, fluorescent materials, invasive ablation, and waste of time. We have studied the CPL scattering technique from both aspects of experimental and computational studies to implement this proposal. To date, we experimentally demonstrated the identification of cancerous parts in sliced biological tissues using the CPL scattering technique with various optical configurations [39]. The line-scanning experiments along a region incorporating normal and cancerous parts show steep changes in the DOCP value depending on the state of the tissue, which indicates the feasibility of this technique in identifying the carcinoma concealed in healthy tissues. Meanwhile, we conducted theoretical and computational analyses using Monte Carlo (MC) simulation methods for the scattering process of CPL with cell nuclei in pseudo-tissues [40]. The single scattering of CPL against particles with diameters corresponding to cell nuclei in healthy and cancerous tissues was investigated first. Then, we introduced them into multiple scattering systems in pseudo-healthy and cancerous tissues to clarify the contribution of optical and structural conditions to the resultant polarization of scattered light. These studies revealed three major points—the resultant DOCP values of the scattered light showed obvious differences between pseudo-healthy and cancerous tissues irrespective of the scattering angles, the intensity of scattered light obtained from healthy and cancerous tissues are approximately the same, and the scattering volume (depth) can be controlled by changing the scattering angle. The third deduction suggests that the depth profile of the tissue can be obtained by analyzing the scattering-angle dependences in the DOCP of the scattered CPL.

In this study, we computationally investigated the quantitative measurement of tumor invasion depth in layered structures that consists of cancerous and healthy tissues. In the preliminary stage, the wavelengths were optimized for detecting cancer and estimating distributions in terms of the polarimetric response for biological tissues and the intensity of scattered light. Subsequently, the changes in the DOCP values were analyzed using the optimized wavelengths for various structure of biological tissues; a cancerous layer lying on the surface [41], and a cancerous layer hiding under the healthy tissue.

2. Experimental methods

The polarization state of light is expressed by the Stokes vector $S$, given by the equation $S = (S_0, S_1, S_2, S_3)^T$, where $S_0, S_1, S_2,$ and $S_3$ are the Stokes polarization parameters [42]. The first Stokes parameter, $S_0$, describes the total intensity of the light beam; the second parameter, $S_1$, describes the preponderance of horizontal LPL over vertical LPL; the third parameter, $S_2$, describes the preponderance of $+45°$ LPL over $-45°$ LPL; $S_3$ describes the preponderance of right-handed CPL over left-handed CPL. The DOCP value, defined by the equation DOCP $= S_3/S_0$, is used to indicate the state of tissues in the CPL scattering technique.

In this study, we used the polarization-light MC algorithm developed by Ramella-Raman *et al.* [43], also known as "meridian plane MC algorithm", to investigate the intensity, polarization, and passage distribution of scattered light. In the polarization-light MC algorithm, light beams are traced using absorption and scattering accompanied by depolarization in a medium. A light beam propagates for a random length associated with the mean free pass in the medium, $(\mu_a + \mu_s)^{-1}$, where $\mu_a$ and $\mu_s$ are the absorption and scattering coefficients of the medium, respectively. The propagation length $\Delta s$ is determined by a random number $\zeta$ ($0 < \zeta \leq 1$), $\Delta s = -\ln(\zeta)/(\mu_a + \mu_s)$. After a traveling in the medium for a distance $\Delta s$, the light beam runs into a scatterer and is subjected to a scattering event. In each scattering process, a random direction and a particular axis are chosen by the rejection method depending on the single scattering phase function in the Mie scattering process [44]. Accordingly, the polarization state after the scattering event is rewritten by the Stokes vector $S$ with respect to the meridian planes, which is determined by the new direction and axis. Simultaneously, a light beam is absorbed at a certain ratio, $albedo = \mu_s/(\mu_a + \mu_s)$. All Stokes parameters are multiplied by $albedo$ to denote a decrease in intensity due to absorption. Subsequently, the light beam travels in a new direction. After this series of processes, the beam going outwards the medium are collectively detected, and the beam whose intensity is less than a certain value falls out as they are fully absorbed. Because of the refraction at the interface to the air, the light beams having an angle less than the critical angle, $\sin^{-1}(1.00/1.33) \approx 48.75°$, can emit an outward air and can be detected, whereas the beams with angles larger than the critical angle are totally reflected and returned to the scattering process loops. These totally reflected light beams scarcely arrive at the surface again, emit outward, and are detected, therefore they are ignored in this study. By repeating these processes from injection to outgoing or full absorption, we analyzed the detected beam for every angle of emergence, hereafter called "detection angle".

Polarized MC simulations were carried out for pseudo-biological tissues with a bi-layered structure: a cancerous layer on a healthy layer and vice versa. The cancerous and healthy tissues are the aqueous dispersions of the particles with 5.9 and 11.0 μm in diameters, respectively. These diameters correspond to the typical nuclear size [21, 45] and the average values we experimentally measured in biological specimens [39]. The refractive indices of the particle and matrix were chosen as 1.59 and 1.33, respectively. The first layered structure, in which a cancerous layer with various thicknesses $t$ is located on a healthy layer, corresponds to cancer that is progressing from the surface to the interior of the tissues. In contrast, the second layered structure, where a cancerous layer lies deep with a depth $d$, corresponds to a buried cancer without exposure to the surface.

## 3. Results and discussion
### 3.1 Optical parameters

The optical parameters, scattering and absorption coefficients, were obtained using semiempirical formulae and experimental measurements obtained from the stomach wall and its carcinoma.

The approximate scattering coefficient $\mu_s'$ as a function of the wavelength is given by Eq. (1) [46, 47].

$$\mu_s'(\lambda) = a \times \lambda^{-b} \text{ (mm}^{-1}) \qquad (1)$$

In this equation, $a$ and $b$ are constants specified by the tissue type. For a stomach wall, $a = 792$ (mm$^{-1}$) and $b = 0.97$ (no units) [48].

Light absorption in a tissue mostly occurs by oxyhemoglobin (HbO$_2$), deoxyhemoglobin (Hb), and water (W). The wavelength dependence of the absorption coefficients ($\mu_a$) is assumed to be approximated by a weighted sum of the spectral absorption coefficients $\mu_{a,\text{HbO}_2}(\lambda)$, $\mu_{a,\text{Hb}}(\lambda)$, and $\mu_{a,\text{W}}(\lambda)$ [46], as given in Eq. (2).

$$\mu_a(\lambda) = S_B\{x \times \mu_{a,\text{Hb}}(\lambda) + (1-x) \times \mu_{a,\text{HbO}_2}(\lambda)\} + S_W \mu_{a,\text{W}}(\lambda) \quad (2)$$

In this equation, $x$ is the oxidation degree of hemoglobin, $x = \text{HbO}_2/(\text{HbO}_2 + \text{Hb})$, and $S_B$ and $S_W$ are heuristic scaling factors adjusted to match the absorption data currently available for each tissue. In ref. [48], the values of these parameters for a normal stomach wall were $x = 0.7$, $S_B = 0.01$, and $S_W = 0.8$. The absorption spectra of these three constituents, $\mu_{a,\text{HbO}_2}(\lambda)$, $\mu_{a,\text{Hb}}(\lambda)$, and $\mu_{a,\text{W}}(\lambda)$, were reported and summarized by Prahl *et al.* [49].

The obtained spectra of the optical parameters were adjusted by the experimental values for stomach wall tissue and stomach tumor tissue. The experimental data were obtained by an integrating sphere and semiconductor lasers of wavelengths 532, 820, and 914 nm. Figure 2 shows the spectra of the optical parameters of healthy stomach wall (red) and gastric cancer (blue), together with the experimental values (black squares with error bars). The experimentally obtained optical parameter values for the healthy tissues were approximately the same as the semiempirical values obtained from Eqn. (1) and (2) for each wavelength. Therefore, we adopted the semiempirical spectra of the optical parameters as those for healthy tissues in this study. The difference in scattering coefficients between the healthy stomach wall and tumor tissue was negligible. As for the absorption coefficient, each experimental value for tumor tissue at three wavelengths was smaller ($-2.7 \sim -3.4$ %) than the values for normal tissues. The average rate of decrease was approximately 3.0 %. Therefore, the spectra obtained by multiplying 0.97 to the spectra obtained by Eqn. (2) were employed as $\mu_a(\lambda)$ spectra for the tumor. These values were then introduced into the algorithms of Monte Carlo simulations to investigate the possibility of cancer detection using the proposed method.

*3.2 Wavelength optimization for cancer detection*

The intensity and polarization of light scattered from pseudo-healthy and cancerous tissues were calculated using the meridian plane MC algorithm. The optical geometry for the calculations is shown in Figure 3(a). The CPL beams whose polarization is $S_3 = +1$ are irradiated into pseudo-biological tissues with an incident angle $\theta = 1°$. The light beams that reflected directly from the surface and the light beams that underwent a few scattering events did not contain important information on the tissue. To exclude these light beams from consideration, the detection region is laid 1 mm from the point of incidence with a width of 1 mm, as shown using the thick red line on the pseudo tissue horizontally 1 to 2 mm in Figure 3(a); only the light beams emitted from this region are detected by CPL detectors faced this region with a detection angle $\varphi$, and included in the calculations. Under this optical configuration, the intensity and DOCP of the light beams detected at the CPL detector were calculated for every detection angle $\varphi$. The $\varphi$ dependences of DOCP and intensity of light at three representative incident wavelengths, 600 nm (red), 750 nm (green), and 950 nm (blue), are shown in Figure 3(b) and (c), respectively. The solid and dashed lines represent the values for healthy and cancerous tissues, respectively. At all wavelengths, the DOCP values of scattered light show a similar (almost parallel) gradual uptrend behavior with respect to $\varphi$. However, the magnitude of DOCP values show significant differences between healthy and cancerous tissues, as well as among three wavelengths. At 600 nm, the difference in the DOCP of light scattered from healthy and cancerous tissues (hereafter, this is defined as $\Delta P = P(\text{healthy}) - P(\text{cancer})$) is negatively large, $\Delta P \cong -0.33$ in wide angular range. In contrast, $\Delta P$ at 950 nm is positively large; that is, the DOCP values obtained from healthy tissue are larger than those from cancerous tissues, which is $\Delta P \cong +0.30$. The intermediate wavelength, 750 nm, shows a small $\Delta P$, $\Delta P \cong +0.01$. The intensities of the scattered light show almost the same angular distribution, with a peak at around 30°. Only the data from healthy tissue at 600 nm showed a slightly small peak, which was derived from the largest absorption coefficient $\mu_a$. To compare $\Delta P$ values among different wavelengths, we employed the average values $\overline{\Delta P}$ within the positive angular range of $\varphi$ (0° ~ 60°) while the sum of intensity $I$ in the same angular range was used for comparison.

Figure 4(a) and (b) show the wavelength dependences of $\overline{\Delta P}$ and $I$, respectively. The average difference in polarization $\overline{\Delta P}$ shows a negative peak at around 600 nm and a positive peak in the near-infrared region from 900 to 1400 nm, together with large vibrations. The total intensity $I$ shows large values within the range of 500 nm to 1300 nm, which is because of the small $\mu_a$. To assess the ability to detect and estimate cancerous tissue depending on the wavelength, the figure of merit $F$ is defined as $F = \overline{\Delta P} \times I$, and its spectral profile is shown in Figure 4(c). In the spectral range less than 500 nm and longer than 1400 nm, significant performances in $\overline{\Delta P}$ are lost owing to the weak intensities. Meanwhile, in the spectral region from green to near infrared, $F$ inherits $\overline{\Delta P}$ behavior, with a negative large peak at around 600 nm, a monotonic increase with large vibrations, and a large positive peak at around 950 nm. The spectral dependence of $F$ clearly indicates that 600 and 950 nm are the appropriate wavelengths for cancer detection, which has a significant difference in polarization with the opposite signs. The largest $F$ derived from the largest $\overline{\Delta P}$ is also obtained at 1050 nm. However, there is a larger fluctuation at around 1050 nm than at 950 nm, due to which this excluded for the optimization. In conventional pulse oximeters, two wavelengths—665 and 880 nm—are used to evaluate the degree of oxygen saturation in the blood. Light at 665 nm has a larger absorbance for hemoglobin than oxyhemoglobin, whereas light at 880 nm has the opposite absorption characteristics [49]. The oxygen saturation is not evaluated by absolute values of absorption but by the ratios between two wavelengths in the pulse oximeters, which ensures the measurement accuracy. Similarly, the accuracy of cancer detection is expected to be enhanced by using the two wavelengths, 600 and 950 nm, which have opposite polarization tendencies for cancerous and healthy tissues.

In this study, we analyzed the reason for the opposite sign in $\overline{\Delta P}$ between the two wavelengths by reviewing a single scattering behavior. Figures 5 shows the calculated results of single scattering for wavelengths $\lambda = 600$ nm (upper row) and 950 nm (lower row). Figure 5(a) and (d) show the scattering angle dependence of intensity $I$ in the polar coordinate with a logarithmic radial axis when a fully right-handed CPL with an angle of incidence of 180° is incident and impinges on a scatterer fixed at the origin (the center of axes). The diameters of scatterer particles are 5.9 $\mu$m (blue: healthy cell) and 11.0 $\mu$m (red: cancerous cell). According to Mie scattering theory [44], the intensities show specific asymmetric behaviors with fine undulations, which is markedly different from the symmetric pattern in the Rayleigh regime in which a scatterer is almost equal to or smaller than the incident wavelength. The number and angular width of the undulation peaks vary depending on the incident wavelength as well as the diameter of the scatterer. In particular, the forward-scattering lobes at around 0°, where the intensity dominantly contributes to the entire scattering, are significantly varied. The intensities at around 0° for a cancerous cell are larger and narrower than those for a healthy cell for both wavelengths. The angular dependences of circular polarization $P$ ($S_3$ in terms of the Stokes parameters), shown in Figure 5(b) and (e), denote complex behaviors with some oscillations, which greatly deviate from the cosine-like shape shown in the Rayleigh regime [21, 40]. To consider the contributions to the resultant polarization, the products of $I$ and $P$ for the wavelengths 600 and 950 nm are plotted in Figure 5(c) and (f), respectively. At 0°, the contributions from the cancerous tissue are larger than those from the healthy tissue, and they are reversed at around 3° at both wavelengths (the insets in Figure 5 (c) and (f)). This indicates that the opposite signs in $\overline{\Delta P}$ between the wavelengths cannot be explained by considering only the contribution within the dominant angular range. The magnitude relations of $\Delta P$ between the wavelengths are not reversed until the contributions of backscattering are considered. In conclusion, the sign of $\overline{\Delta P}$ shown in Figure 4 depends largely on the intensity and polarization of the backscattered light. The contributions of the backscattered light are extremely complicated and difficult to define universally. Moreover, the calculation results shown in Figure 4 cannot be applied to a transparent configuration that is employed, for example, in a fingertip pulse oximeter, because the forward scatterings are used dominantly.

*3.3 Depth estimation of tumor invasion*

The measurements of the depth profiles were verified using the two optimized wavelengths. An example of the distribution of simulated light beam paths is shown in the area of a biological tissue of Figure 3(a), which is obtained under the conditions that the detection angle is $\varphi = 35 \pm 5°$, the wavelength $\lambda = 950$ nm, the medium comprises spheres of diameter $a = 5.9$ µm, (a pseudo-healthy tissue), and a photon number of the incident CPL is 500,000. The simulated light beam paths under the other condition ($\lambda$ and $a$) are shown in Figure S1 to S4 in the Supporting Information. The distribution of light beams drastically varied with $\varphi$: changing from $-90°$ to $+90°$ while maintaining $\theta = 1°$. When $\varphi$ is close to zero (vertical incidence), the detected light beams contain many light beams that dive deeper. In contrast, light beams scattered in a shallow volume tend to exit from the surface with a large $\varphi$. The averaged maximum depth of the detected light beams is defined here as $L$, which can be paraphrased as a sampling depth. Figure 6 shows $L$ values as functions of $\varphi$. The values of $L$ increase monotonically with $\varphi$ in the whole angular range irrespective of wavelengths and tissue states, which indicates that the sampling depth can be modulated by tuning $\varphi$. Therefore, the depth profile can be obtained from the detection angle dependence of the DOCP values.

By taking advantage of the tunable sampling depth, the cancer distributions along the depth direction were examined with MC simulations. The schematic representations of the optical configurations and the structure of pseudo-biotissues are shown in Figure 7(a) and (c). As shown in Figure 7(a), the pseudo-biotissues having a cancer layer with the thickness of $t$ on a healthy layer represent cancer tissues progressing deeper from the surface, which is used for measuring the tumor depth invasion in the early stages of cancer. Conversely, the buried cancer layer shown in Figure 7(c) corresponds to the pseudo-tissues that lies hidden beneath a healthy layer with a depth of $d$. This cancer layer is assumed to be an intraepidermal carcinoma concealed with epithelial tissues or the tissue at the marginal region of cancer with invasive spreading into the submucosa layer. The calculated DOCP values for 600 and 950 nm as a function of structural parameters, $t$ and $d$, are shown in Figure 7 (b) and (d), respectively. (The $\varphi$ dependences of intensity and DOCP values are shown in Figure S5 and S6 in the Supporting Information.) The variation of DOCP with $t$ and $d$ show an opposite tendency between 600 and 950 nm wavelengths, which is attributed to the opposite sign in $\overline{\Delta P}$. For superficial cancer, when $t$ increases from 0 to approximately 1.0 mm, the DOCP values increase and decrease monotonically for 600 and 950 nm, respectively. In this thickness region, most of the light beams reach the underlaid healthy layers. Therefore, the DOCP values change depending on the ratio of the cancerous layer volume to the entire sampling volume. When $t$ increases further, there are two types of DOCP behavior depending on the detection angle $\varphi$: the DOCP values for large $\varphi$ are saturated at 1.4 mm, while the values for small $\varphi$ continue to change up to 2.0 mm. The difference in the DOCP values between $\varphi$ becomes small at 600 nm and large in 950 nm. This difference in DOCP values is due to the fact that the scattering volume of light with large $\varphi$ is fulfilled with a cancer layer, whereas the scattering volume of light with small $\varphi$ still includes healthy tissues. A further increase in $t$ ($t \geq 2.0$ mm) induces saturation of DOCP values in the entire angular range, because the scattering volumes are fulfilled with a cancer layer. Taken together, the depth profile can be obtained by comparing the DOCP values of a healthy part ($t = 0$) and of a target point for a cancer thinner than 1.0 mm, and by comparing of DOCP values of different $\varphi$ for a cancer thicker than 1.0 mm and up to approximately 1.8 mm. Under existing conditions, the detection limit for the quantitative measurement of cancer thickness is approximately 2.0 mm, which is sufficient to diagnosis whether the cancers stay in the mucosa whose thickness is approximately 1.0 mm (Tis, T1a, and T1b in Figure 1) or progress beyond the mucosa to the underneath layers (Ts or worse in Figure 1). Due to the opposite tendencies between the two wavelengths, the difference in DOCP values between $\varphi = 0°$ (purple) and $60°$ (red) in the thickness range $t > 1.0$ nm decreases for $\lambda = 600$ nm and increases for $\lambda = 950$ nm. Such variations can be used for a more accurate estimation of cancer thickness.

On the other hand, the calculation results for a buried cancerous layer beneath a healthy layer shows the opposite behavior to the results for the superficial cancer at 600 and 950 nm (Figure 7(d)). However, similar to the superficial cancer, the comparison between the DOCP values of a healthy tissue and a target point can provide the cancer depth for a cancerous layer lying in a shallow place, while the depth of cancer lurking in the deeper can be evaluated from the difference in DOCP values between different $\varphi$. For 600 nm, the differences in the DOCP values between different $\varphi$ are small within the depth range, $0 < d < 1.0$ mm. This characteristic of 600 nm light can be used for a more accurate depth estimation of cancer tissues buried shallower than 1.0 mm. In contrast, the depth measurement with CPL at 900 nm is possible in the wider $d$ range, $0 < d < 1.6$, because the DOCP values continuously change to the deeper region. The depth detection limit is approximately 1.6 mm for this method. Two wavelengths exhibiting the opposite responses from healthy and cancerous tissues enable us to distinguish between the cancer on the surface and buried in the depth, as well as to extract the valid signals from noise to increase the reliability.

## 4. Conclusions

We conducted a computational analysis of the quantitative depth estimation of tumor invasion in early gastric cancer using the CPL scattering technique. First, the optical parameters, the scattering and absorption coefficients, of the human stomach wall and its carcinoma were obtained by semiempirical and experimental methods. By introducing the obtained parameters into MC algorithms, the differences in circular polarization, $\overline{\Delta P}$, and intensity, $I$, of the resultant scattered light were calculated for a wide range of wavelength: from visible to near-infrared light. The wavelength dependence of the figure of merit, $F =$ $\overline{\Delta P} \times I$, indicates that 600 and 950 nm have considerable differences in circular polarization with the opposite signs and large intensity, indicating their appropriateness for cancer detection. The opposite signs of $F$ at 600 and 950 nm emerge from the contribution of the backscattered light. At the two optimized wavelengths, the sampling depth in the biological tissues can substantially depends on the detection angles, indicating that the depth profile can be detected by tuning the detection angle. The calculated results for the pseudo-biological tissues of the bi-layered structure containing a cancerous and healthy layer indicate the successive changes in the DOCP values depending on the thickness or the depth of cancerous layer. The structure consisting of a cancerous layer on a healthy layer corresponds to an early-stage caner, in which cancer progresses deeper from the surface. The cancer thickness in this structure can be evaluated by comparing the DOCP values in a completely healthy tissue and a progressing cancer tissue for thin cancer. For the further progressed cancer (thicker cancer), the difference in DOCP values between the different $\varphi$ allowed us to determine the thickness. In the cancerous layer crept under the healthy layer, the cancer depth can be estimated from the same comparisons. The two wavelengths exhibiting the opposite tendency against the thickness and depth of the cancerous layer can not only increase the accuracy of estimation but also facilitate noise elimination from the detected polarization signal, that is, offsets with the same tendency between two wavelengths in the detected signals can be inferred to be signals from unintended factors and can be eliminated as noise. The identification limit of the thickness and the depth of the estimated cancer is approximately 2.0 mm and 1.6 mm, respectively. The thickness resolution notably depends on the detection resolution of the DOCP values in the CPL detectors. At least 0.1 or less resolution of DOCP values is sufficient to discriminate cancer remaining in the mucosa (Tis or T1) from the cancer progressed deeper than the submucosa (T2 or the further), which will provide quantitative information effective for diagnosis concerning the therapeutic approach, endoscopic treatment or surgical procedure with dissections. The simultaneous detection of DOCP values with different detection angles is possible with an endoscopic probe consisting of a CPL emitter, some CPL detectors, and a parabolic mirror attached to the tip of an endoscope, which we proposed in Ref. [40]. However, the tip of the endoscope is crowded because of the recent multi-functionalization of endoscopes. Alternatively, we propose to attach the endoscopic probe to the sidewall of an endoscope, which we obtained from the intravascular optical coherence tomography [50, 51]. The smaller spatial restrictions will enable the longer semi-latus rectum of a parabolic mirror to detect light with a detection angle within a large angular range. To estimate the depth of invasion tumors in gastric cancer using the CPL scattering technique, experimental demonstrations using cancerous biological specimens at various stages of early gastric cancer are required. Moreover, further developments are needed in the CPL emitter and detector based on spin-LEDs used for this technique.

**Ethics approval**

Ethics approval is not required to carry out this work.

**AUTHOR CONTRIBUTIONS**

N. N. was involved in conceptualization, investigation, calculations, writing—original draft preparation, writing—review and editing, project management. T. K. prepared the biological tissues and was involved in review and editing from the biological viewpoint.

**CONFLICT OF INTEREST**

The authors declare no potential conflict of interests.

**ACKNOWLEDGEMENTS**

This work was partially supported by KAKENHI (Nos. 17K14104, 18H03878, and 19H04441) of the Japan Society for Promotion of Science (JSPS), the Cooperative Research Project of Research Center for Biomedical Engineering, Futaba Foundation, the Uehara Memorial Foundation, and a Grant-in-Aid for Challenging Research, Organization of Fundamental Research, Tokyo Institute of Technology (TIT). The authors acknowledge Prof. Hiro Munekata for fruitful discussions and Mr. Shinya Kawashima for technical support at TIT.

**DATA AVAILABILITY**

The data that support the findings of this study are available from the corresponding author upon reasonable request.

**Figure Captions**

**Fig. 1** Clinical stages of gastric cancer. Tis (Carcinoma *in situ*; intraepithelial tumor without invasion of the lamina propria), T1 (Tumor invades the lamina propria or muscularis mucosae (T1a), and submucosa(T1b)), T2 (Tumor invades the muscularis propria), T3 (Tumor penetrates the subserosal connective tissue), and T4 (Tumor invades the serosa (T4a) or adjacent structures (T4b)) [2-4]

**Fig. 2** Wavelength dependence of the optical parameters, scattering coefficient $\mu'_S$ and absorption coefficient $\mu_a$, of the healthy stomach wall (red) and wall with gastric cancer (blue), respectively, together with the experimental values (black squares with error bars).

**Fig. 3** (a) Monte Carlo simulation geometry for multiple scattering in pseudo biological tissues together with the calculated distribution of light beam paths under the condition that $\varphi = 35 \pm 5°$, $a = 5.9$ μm, and $\lambda = 950$ nm. The detection angle dependence of the calculated (b) DOCP and (c) intensity for $\lambda = 600$ nm (red), 750 nm (green), and 950 nm (blue). The solid and dashed lines represent the values for healthy and cancerous tissues, respectively.

**Fig. 4** Wavelength dependence of (a) $\Delta P$, (b) $I$, and (c) figure of merit $F$. (a) $\Delta P$ values are obtained from the averaged difference in polarization of the scattered light, where $\Delta P = P(\text{healthy tissue}) - P(\text{cancerous tissue})$ within the positive angular range of $\varphi$ (0° ~ 60°). (b) $I$ is sum of intensity of the scattered light in the same range. (c) $F$ is obtained as the product of $\Delta P \times I$.

**Fig. 5** Calculation results for single scattering against a particle with $\lambda = 600$ nm on the upper row and 900 nm on the lower row, respectively. The red and blue lines show the results for a cell nucleus $a$ in cancerous ($a = 11.0$ μm) and healthy tissues ($a = 5.9$ μm), respectively. The repetition number is 100,000 (a) (d) Intensity of scattered light $I$ as a function of scattering angles when a light beam ($S_3 = +1$) comes with an angle of incidence 180° and impinged on a particle at the origin. The calculated intensity values are plotted on a logarithmic radial axis. (b) (e) DOCP value $P$ as a function of scattering angles. (c) (f)

Angle dependences of the product of $I \times P$ which are the relative expectation values of DOCP values for single scattering. The insets show the magnified graph in the small angular region (15° ~ 0°).

**Fig. 6** The detection angle $\varphi$ dependence of sampling depth $L$ for pseudo-healthy tissue ($a = 11.0$ μm: blue plots and lines) and pseudo cancerous tissue ($a = 5.9$ μm: red plots and lines) with $\lambda = 600$ nm (closed squares) and $\lambda = 950$ nm (opened squares).

**Fig. 7** The schematic optical geometry and the layered structures of the pseudo-tissues for (a) a cancerous layer lying on the surface progresses deeper and (c) cancer hiding under the healthy tissue. The calculated resultant DOCP values of light scattered from pseudo-tissues as a function of (b) thickness of cancer $t$ and (d) depth of cancer $d$ with different detection angle $\varphi$ for $\lambda = 950$ nm (opened squares) and $\lambda = 600$ nm (closed squares).

Figure 1

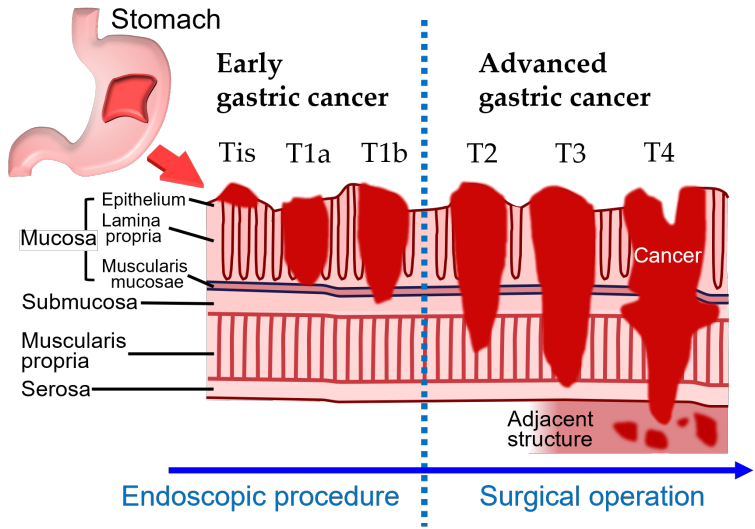

N. Nishizawa *et al.*,

Figure 2

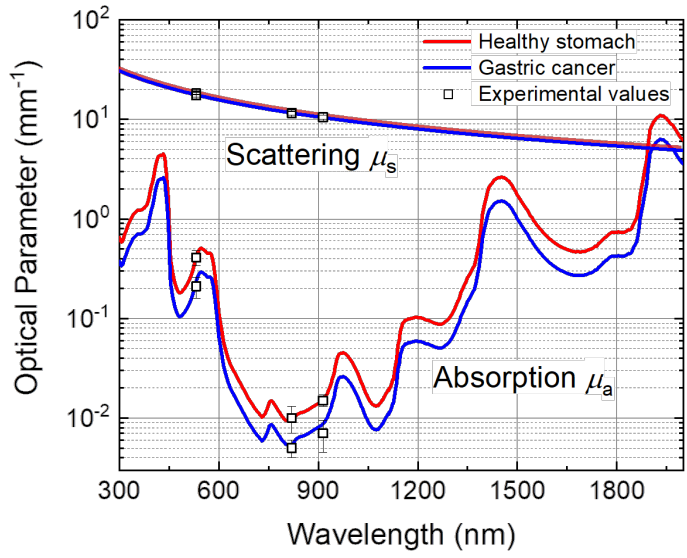

N. Nishizawa *et al.*,

Figure 3

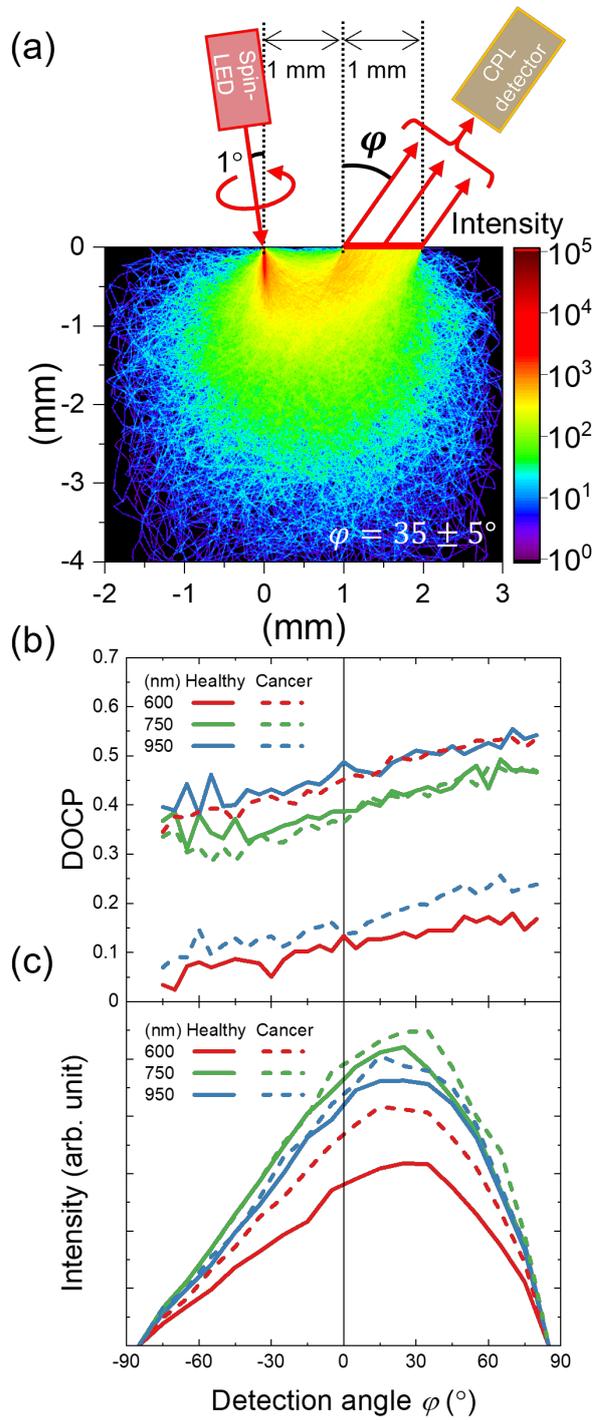

N. Nishizawa *et al.*,

Figure 4

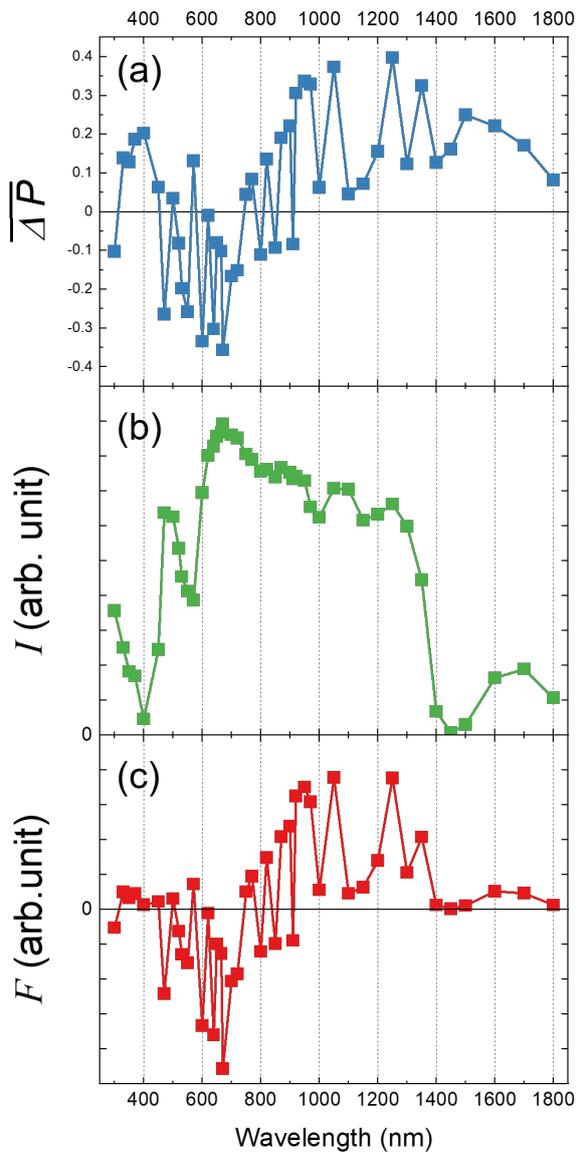

N. Nishizawa *et al.*,

Figure 5

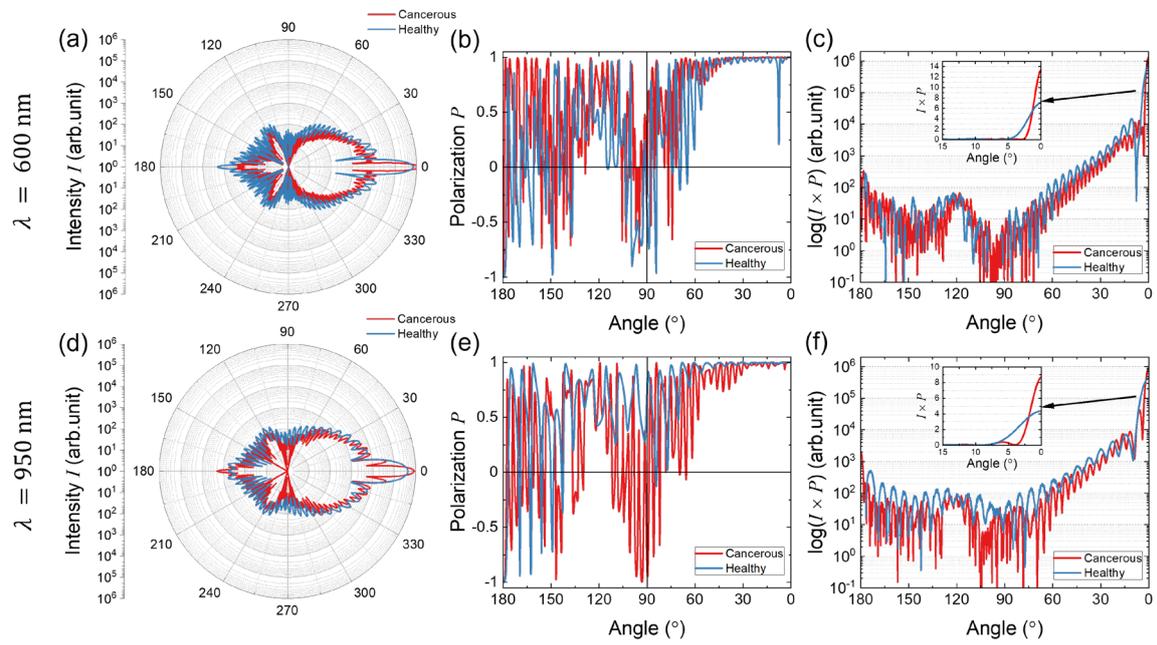

N. Nishizawa *et al.,*

Figure 6

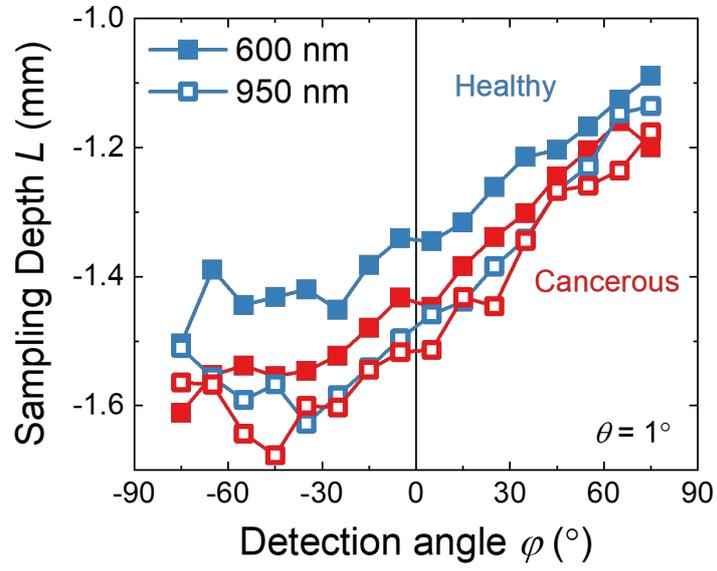

N. Nishizawa *et al.*,

Figure 7

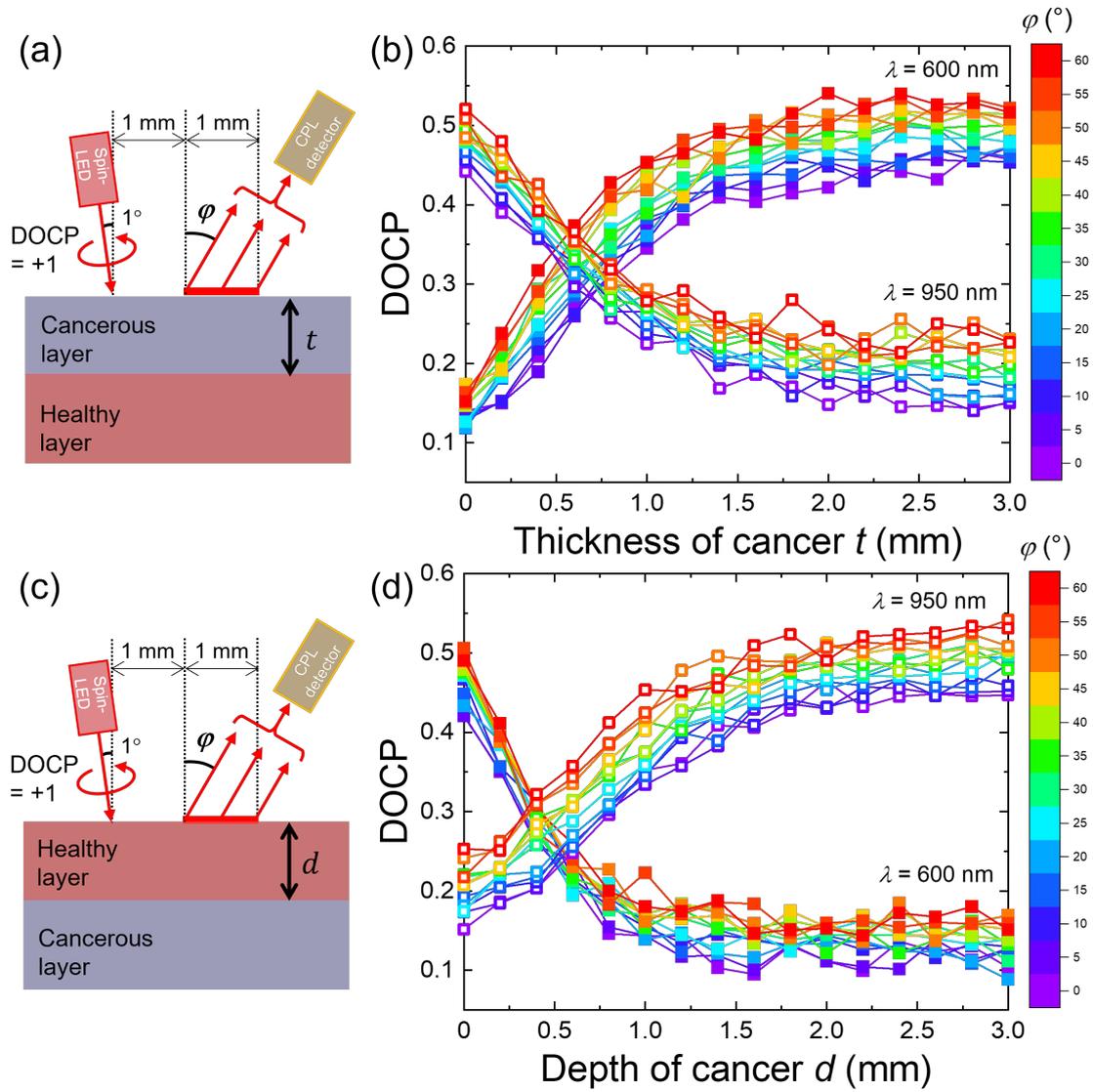

N. Nishizawa *et al.,*

# Supporting Information

# Depth estimation of tumor invasion in early gastric cancer using scattering of circularly polarized light: Monte Carlo Simulation study

Nozomi Nishizawa, and Takahiro Kuchimaru

**Supplementary Data 1: Light paths**

Figure S1–S4 show the distribution of simulated light beam paths of wavelength $\lambda$ for the pseudo-biotissues comprising spheres of diameter $a$. Figure S1, S2, S3, and S4 are the results under the conditions $(\lambda, a) = (600\text{ nm}, 11.0\text{ μm}), (600\text{ nm}, 5.9\text{ μm}), (950\text{ nm}, 11.0\text{ μm})$, and $(950\text{ nm}, 5.9\text{ μm})$, respectively, which include 16 figures for different $\varphi$ from $-75°$ to $+75°$ with each $10°$ angular width. The repetiation (photon) numbers are $N = 500,000$. The medium having spheres of diameter $a = 5.9\text{ μm}$ and $11.0\text{ μm}$ corresponds to the pseudo-healthy and cancerous tissues, respectively. The 12th figure in Figure S4 is identical with Figure 3 (a) in the main text.

**Supplementary Data 2: Intensity and DOCP values in the layered structure**

Figure S5 and S6 show the detection angle $\varphi$ dependences of intensity and DOCP in the layered structure. These are the results for a cancerous layer lying on the surface progresses deeper (Figure S5) and cancer hiding under the healthy tissue (Figure S6). These data are shown in Figure 7 in the main text as a function of the structure parameters. The detection angle dependence of the calculated (a)(c) intensity and (b)(d) DOCP values for (a)(b) $\lambda = 600\text{ nm}$ and (c)(d) 950 nm, respectively. The colors of the plot represent the thickness of cancer $t$ (Figure S5) and the depth of cancer $d$ (Figure S6). Figure 7 (b) and (d) correspond $t$ and $d$ dependences of DOCP values shown in Figure S5(b)(d) and S6(b)(d), respectively.

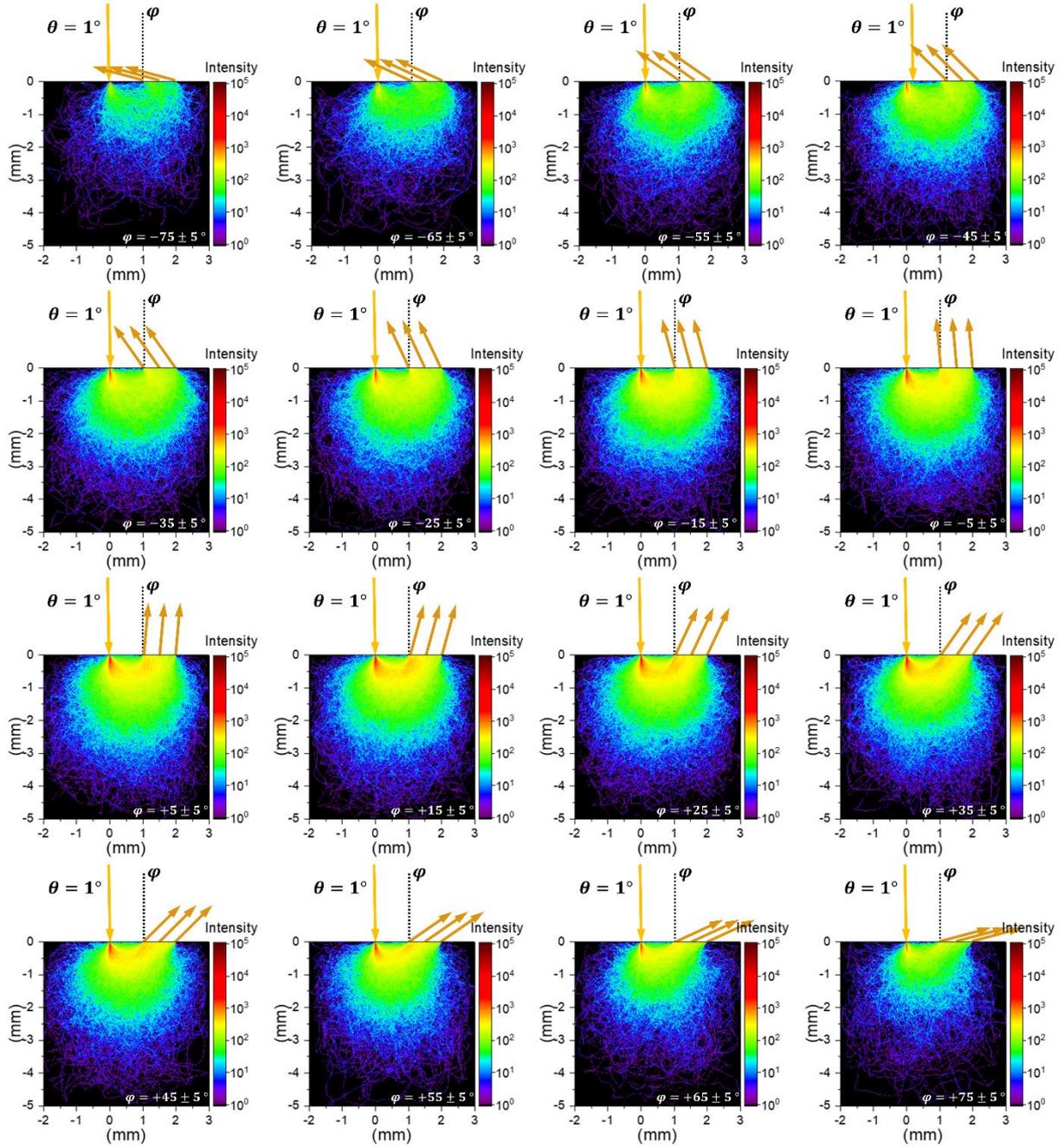

**FIGURE S1** The distribution of simulated light beam paths under the condition that $\lambda = 600$ nm, $a = 11.0$ μm, $N = 500{,}000$, and $\varphi = -75° \sim +75°$.

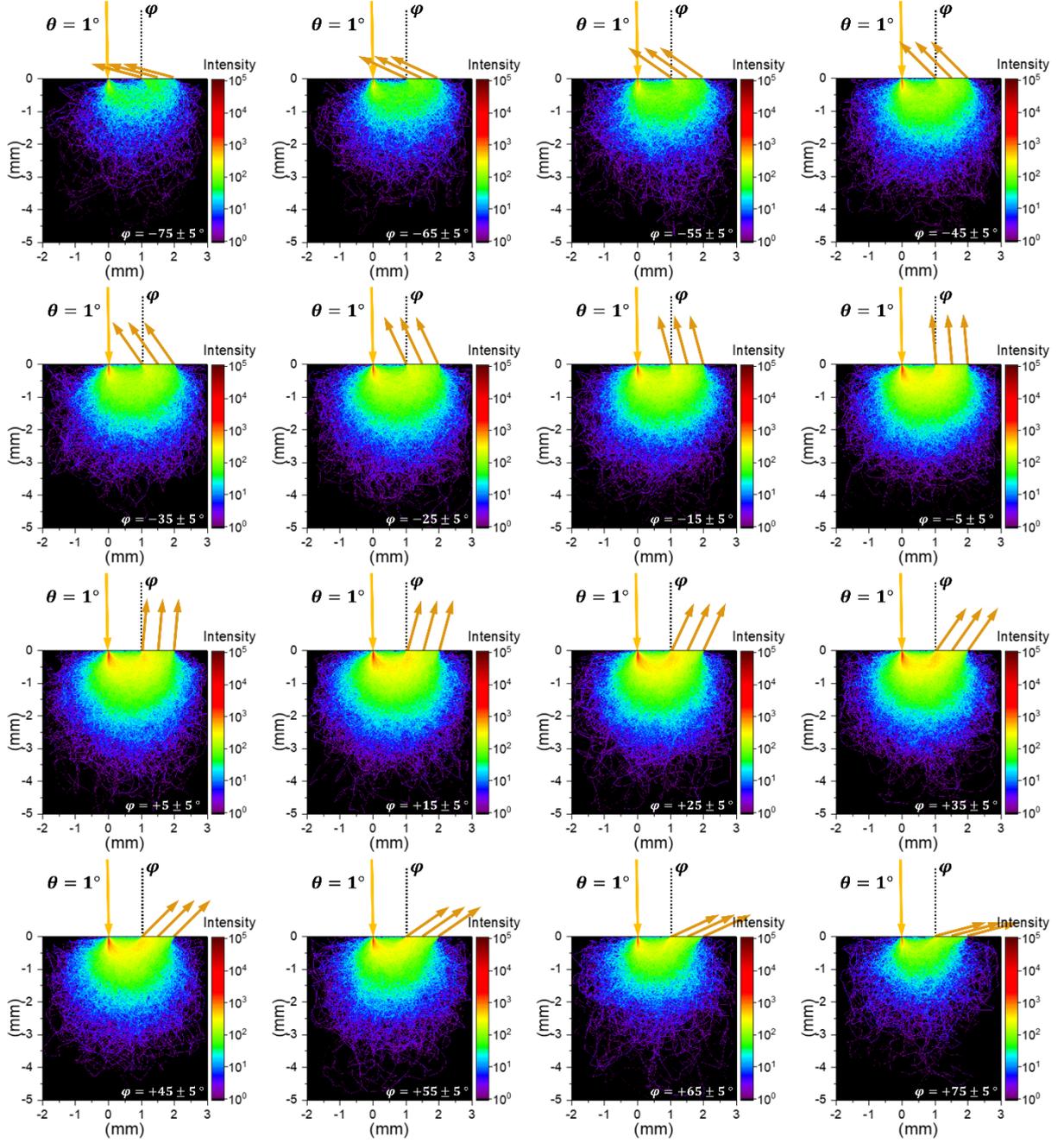

**FIGURE S2** The distribution of simulated light beam paths under the condition that $\lambda = 600$ nm, $a = 5.9$ μm, $N = 500{,}000$, and $\varphi = -75° \sim +75°$.

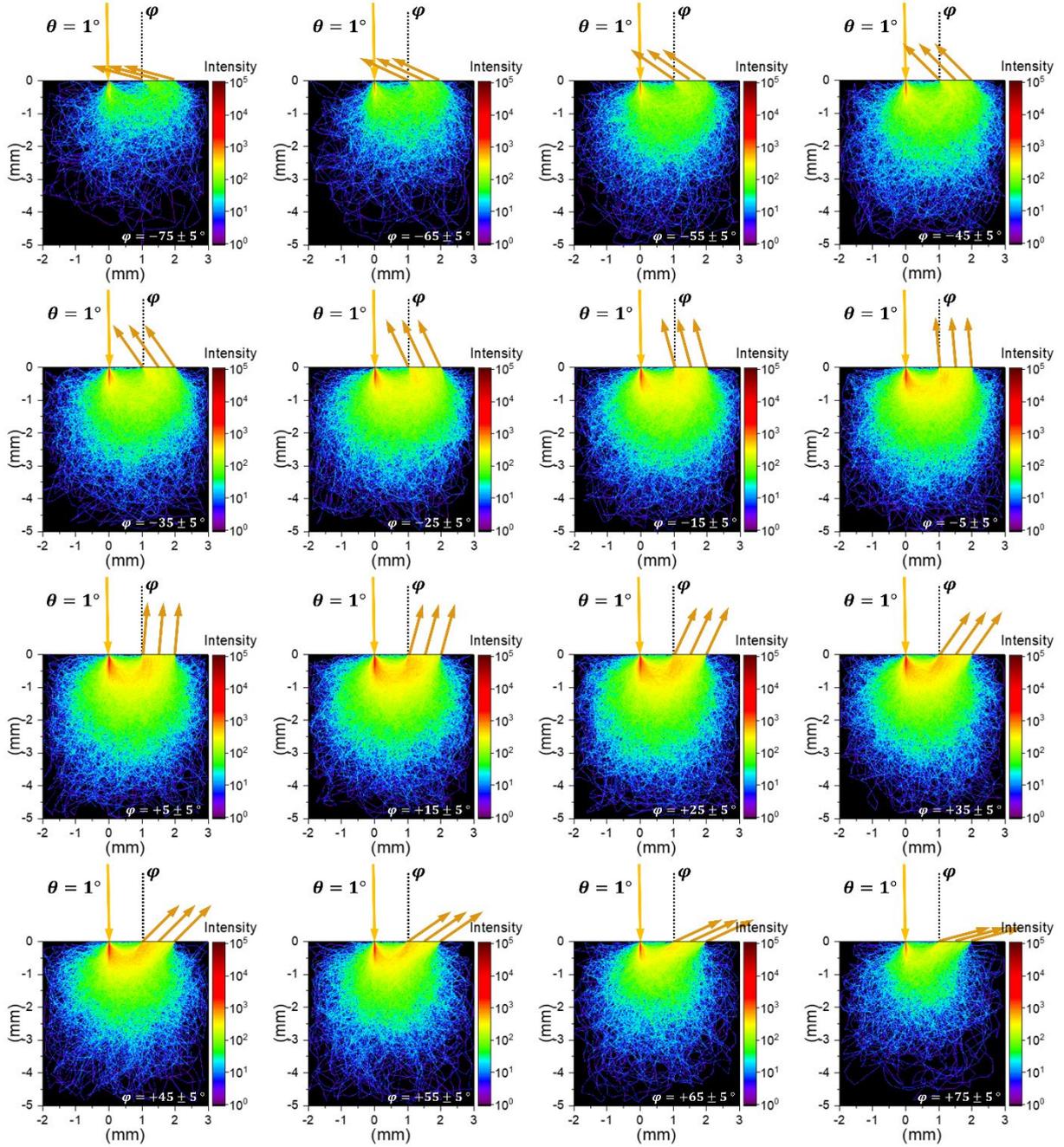

**FIGURE S3**  The distribution of simulated light beam paths under the condition that $\lambda = 950$ nm, $a = 11.0$ μm, $N = 500{,}000$, and $\varphi = -75° \sim +75°$.

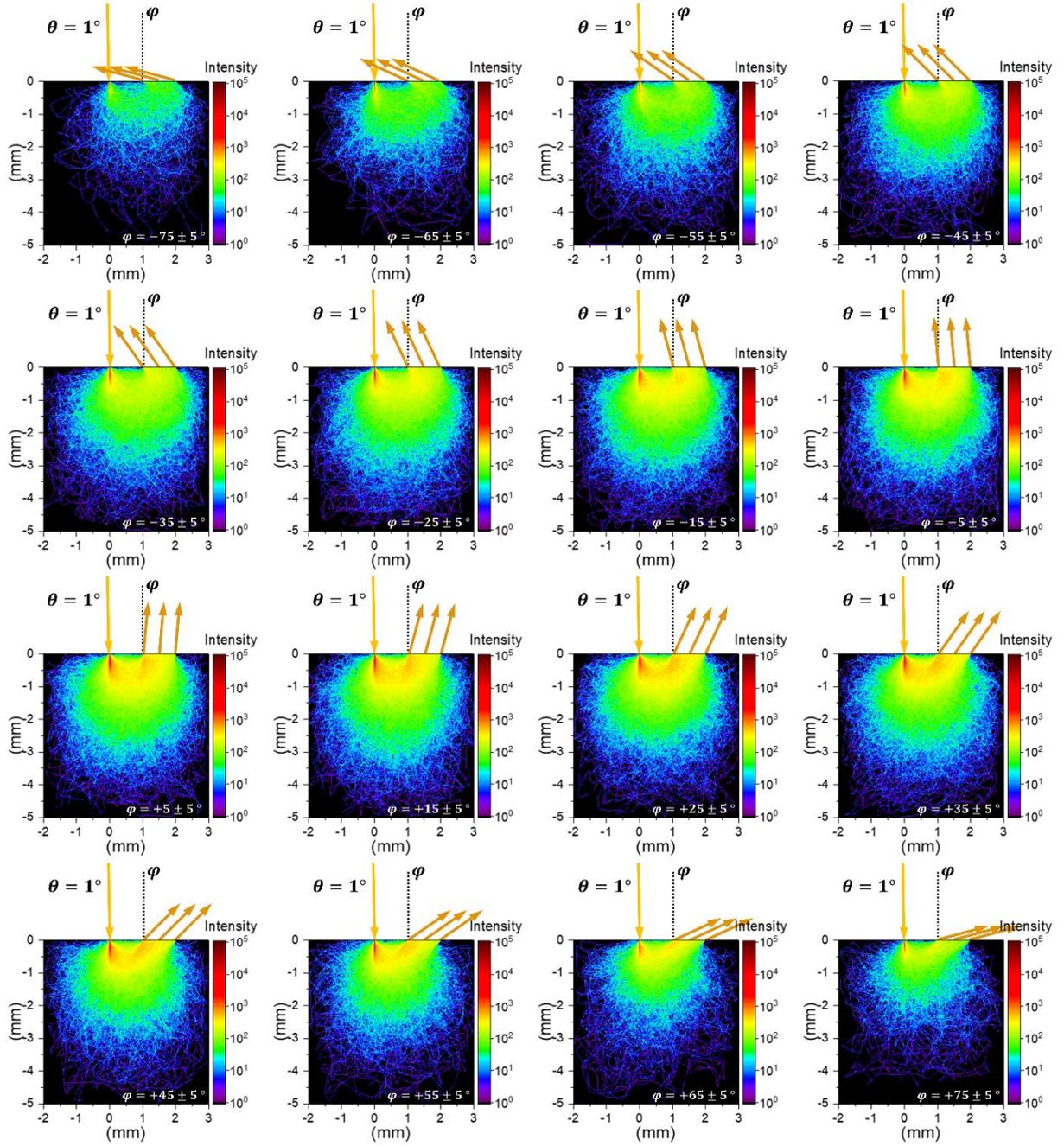

**FIGURE S4** The distribution of simulated light beam paths under the condition that $\lambda = 950$ nm, $a = 5.9$ μm, $N = 500{,}000$, and $\varphi = -75° \sim +75°$.

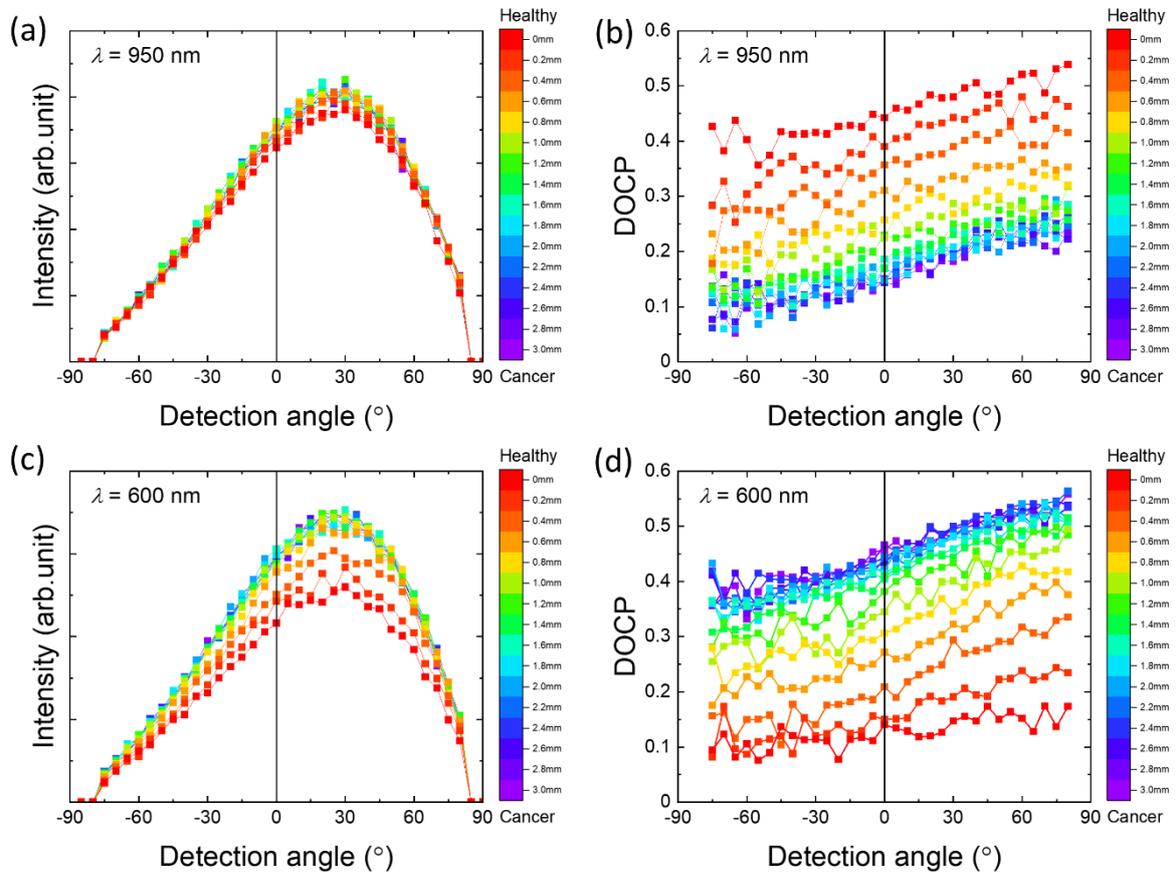

**FIGURE S5** The detection angle dependence of the calculated (a)(c) intensity and (b)(d)DOCP values for (a)(b) $\lambda = $ 600 nm and (c)(d) 950 nm, respectively. The colors of the plot represents the thickness of cancer $t$.

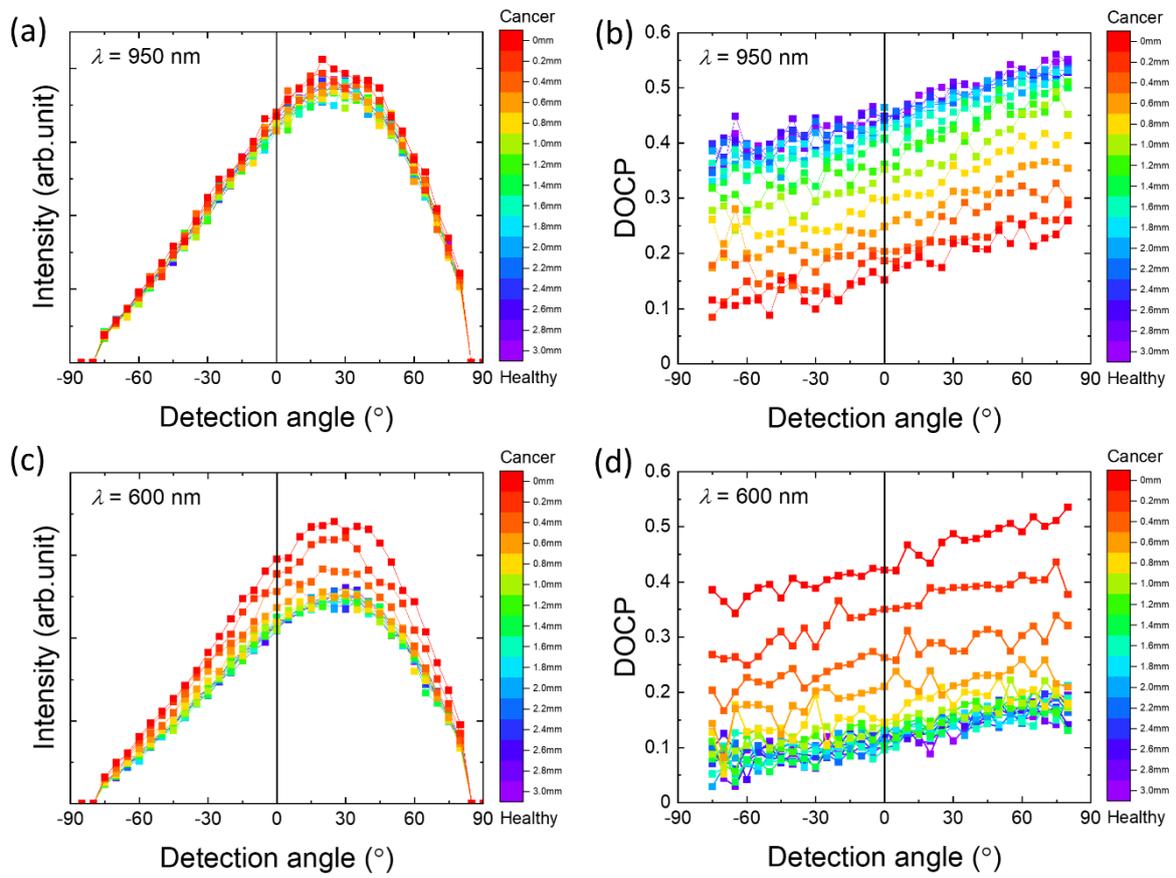

**FIGURE S6** The detection angle dependence of the calculated (a)(c) intensity and (b)(d) DOCP values for (a)(b) $\lambda = $ 600 nm and (c)(d) 950 nm, respectively. The colors of the plot represents the depth of cancer $d$.